\documentclass{elsart}

\usepackage{natbib}

\def\p{{\rm peak}}
\def\iso{{\rm iso}}

\def\sw{{\em Swift}}

\def\spose#1{\hbox to 0pt{#1\hss}}
\newcommand\lsim{\mathrel{\spose{\lower 3pt\hbox{$\mathchar"218$}}
     \raise 2.0pt\hbox{$\mathchar"13C$}}}
\newcommand\gsim{\mathrel{\spose{\lower 3pt\hbox{$\mathchar"218$}}
     \raise 2.0pt\hbox{$\mathchar"13E$}}}

\usepackage{amssymb}
\usepackage{graphicx}

\begin{document}

\begin{frontmatter}



\title{Are Gamma-Ray Bursts Standard Candles?}


\author{Li-Xin Li}

\address{Max-Planck-Institut f\"ur Astrophysik, 85741 Garching, Germany}

\begin{abstract}
By dividing a sample of 48 long-duration gamma-ray bursts (GRBs) into four 
groups with redshift from low to high and fitting each group with the Amati 
relation $\log E_\iso = a + b \log E_\p$, I find that parameters $a$ and $b$ 
vary with the mean redshift of the GRBs in each group systematically and 
significantly. The results suggest that GRBs evolve strongly with the cosmic 
redshift and hence are not standard candles. 
\end{abstract}

\end{frontmatter}


A remarkable achievement in the observation of GRBs has
been the identification of several good correlations among GRB
observables [see \citet{sch07} for a review]. Based on several of those
correlations, some people have eagerly proposed that GRBs are standard
candles and can be used to probe cosmology to very high redshift through 
the Hubble diagram \citep{sch03,dai04,ghi04,lam05,fir06}.

Type Ia supernovae (SNe Ia) have been considered as standard candles and 
applied to cosmology, which has produced a persuasive evidence that the 
Universe is currently expanding with an accelerating speed 
\citep[and references therein]{clo06}. 

However, the situation of GRBs is
very different from that of SNe Ia. The majority of the observed SNe Ia 
have redshift $z\lsim 0.1$, so the evolution of SNe Ia plays a minor role. 
While for GRBs, all the correlations have been obtained by fitting a 
{\em hybrid} GRB sample with redshift spanning a very large range: from 
$z\sim 0.1$ up to $z \sim 6$. For objects in such a large range of redshift, 
it is hard to believe that evolution is not important.

In addition, the physics of SNe Ia is much better understood than that of 
GRBs. We know that SNe Ia are produced by the thermal nuclear explosion
of white dwarfs. For GRBs we are not aware of their progenitors, although 
many people think that long-duration GRBs arise from the core-collapse of
rapidly rotating massive stars \citep[and references therein]{pir04}.

The fact that people use a hybrid GRB sample to do statistics without 
discriminating the redshift distribution is of course caused by the fact 
that we do not 
have an enough number of GRBs with measured redshifts limited in a small 
range. Then, inevitably, the effect of the GRB evolution with redshift, 
and the selection effects, have been ignored. This raises an important 
question about whether the relations that people have found reflect the 
true physics of GRBs or they are superficial \citep{ban05}. 

A necessary condition for a class of objects to be standard candles is that 
they do not evolve with the cosmic redshift. Or, they evolve with the redshift 
but we know how they evolve (V. Petrosian, this proceeding). For GRBs, neither
of these conditions is satisfied.

I use the Amati relation as an example to test the 
cosmic evolution of GRBs. The Amati relation is a correlation between the
isotropic-equivalent energy of long GRBs and the peak energy of 
their integrated spectra in the GRB frame \citep{ama02}
\begin{eqnarray}
        \log E_\iso = a + b \log E_\p \;. \label{amati}
\end{eqnarray}
With 41 long GRBs with firmly determined redshifts and peak 
spectral energy, \citet{ama06} has obtained that $a= -3.35$ and $b= 1.75$ 
with the least squares method ($E_\p$ in keV and $E_\iso$ in $10^{52}$ erg).

If GRBs do not evolve with the cosmic redshift, we would expect that the 
Amati relation in equation (\ref{amati}) does not change with the redshift. 
Hence, a test on the variation of the Amati relation with the redshift 
could provide some constraint on the cosmic evolution of GRBs.

For this purpose, I separate a sample of 48 long GRBs [41 GRBs from 
\citet{ama06}, and seven additional \sw\, GRBs from \citet{ama07}] into four 
groups by the redshift, each group containing 12 GRBs: {\em Group A} -- $0.1< 
z <0.84$, $\langle z\rangle = 0.56$; {\em Group B} -- $0.84\le z <1.3$, 
$\langle z\rangle = 1.02$; {\em Group C} -- $1.3\le z <2.3$, $\langle 
z\rangle = 1.76$; {\em Group D} -- $2.3\le z \le 5.6$, $\langle z\rangle = 
3.40$; where $\langle z\rangle$ is the mean redshift.

\begin{figure}
\includegraphics[angle=0,scale=0.375]{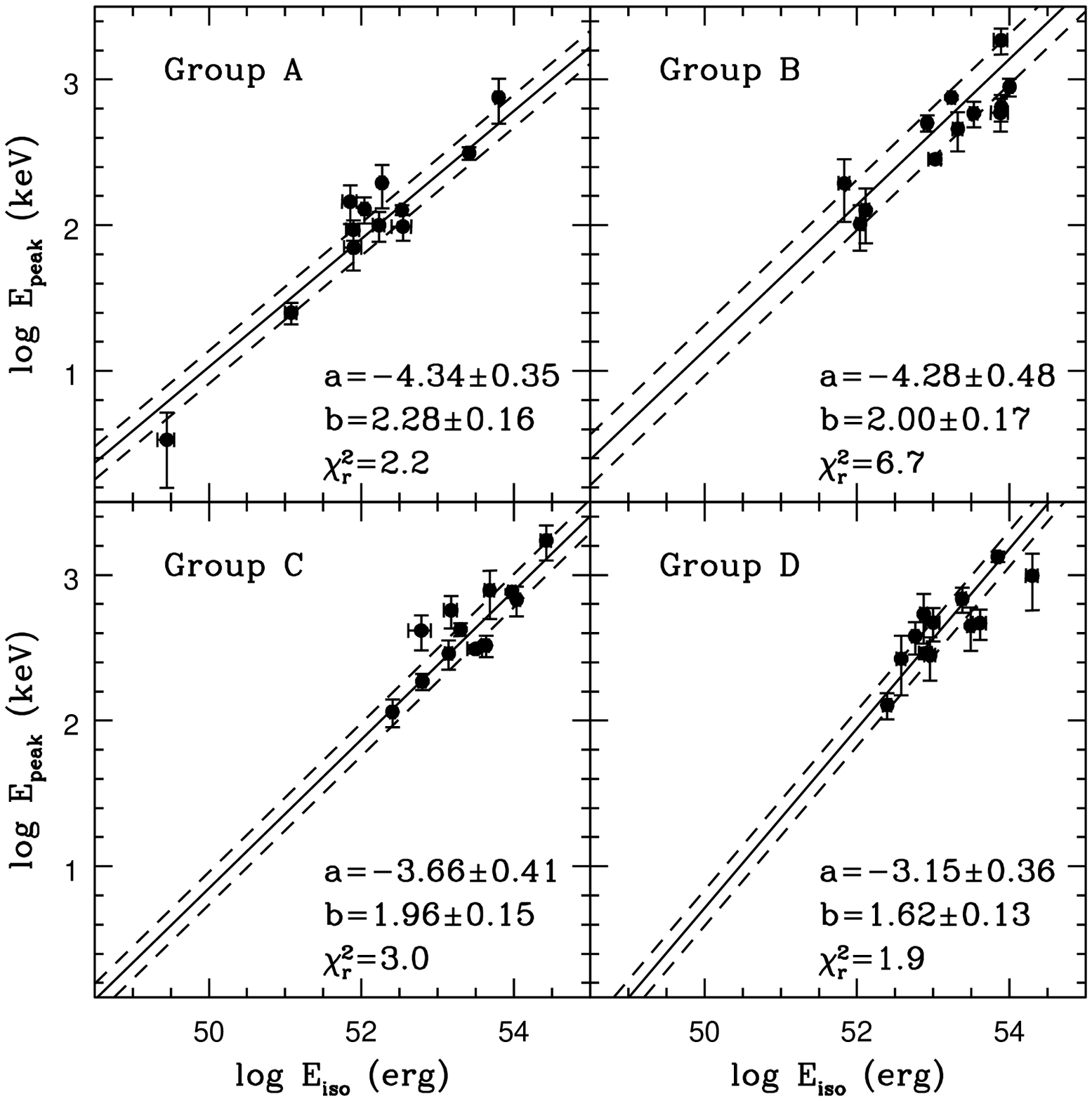} \hspace{0.47cm}
\includegraphics[angle=0,scale=0.358]{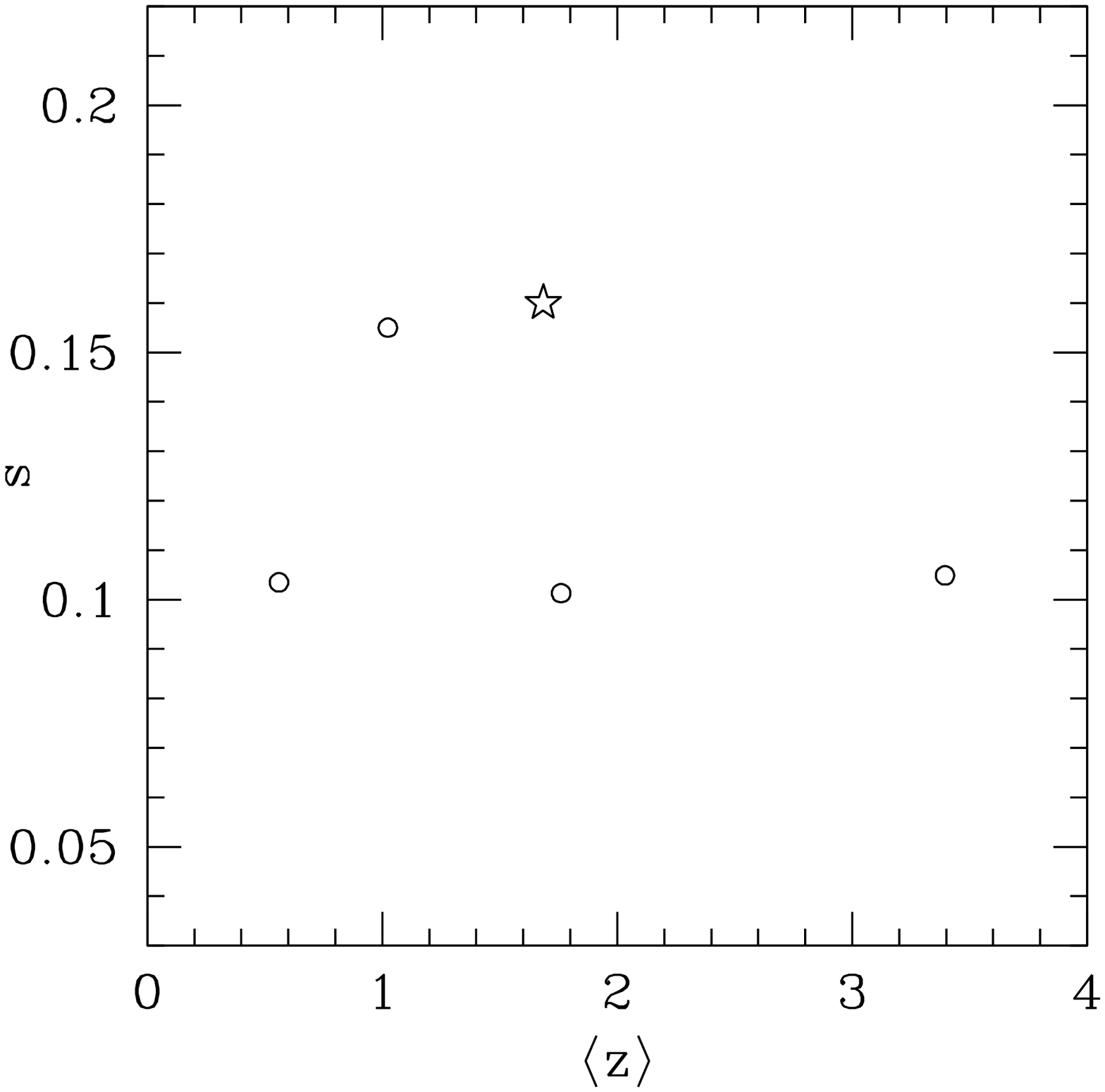}
\caption{Left panel: Least squares fit to each of the four groups of GRBs by 
equation (\ref{amati}) (solid line). The two dashed lines mark the 1-$\sigma$ 
deviation of the fit. Right panel: The deviation of fit. Each circle 
corresponds to a group of GRBs. The star ($s=0.16$) is the result obtained by 
fitting the whole sample (48 GRBs).
}
\label{ee_z2}
\end{figure}

Then, I fit each group by equation (\ref{amati}) and calculate the mean 
redshift, and check if the values of $a$ and $b$ evolve with the redshift.
The results are shown in Fig.~\ref{ee_z2} (left panel; $\Omega_m = 0.3$, 
$\Omega_\Lambda=0.7$, and $H_0 = 70$ km s$^{-1}$ Mpc$^{-1}$).

A least squares fit to the 48 GRBs as a single sample with equation 
(\ref{amati}) leads to $a= -3.42$ and $b= 1.78$, with the reduced $\chi_r^2 = 
5.9$. The results are consistent with that obtained with 41 GRBs by 
\citet{ama06}. In the right panel of Fig.~\ref{ee_z2}, I plot the deviation 
of fit [$s$; see \citet{li06}] against the mean redshift of GRBs. There is 
not a clear trend for $s$ to vary with $\langle z\rangle$. But it appears 
that the deviation of fit of each group is smaller than that of the whole 
sample. This fact indicates that treating GRBs at different redshifts as a 
single sample may increase the data dispersion.

\begin{figure}
\includegraphics[angle=0,scale=0.35]{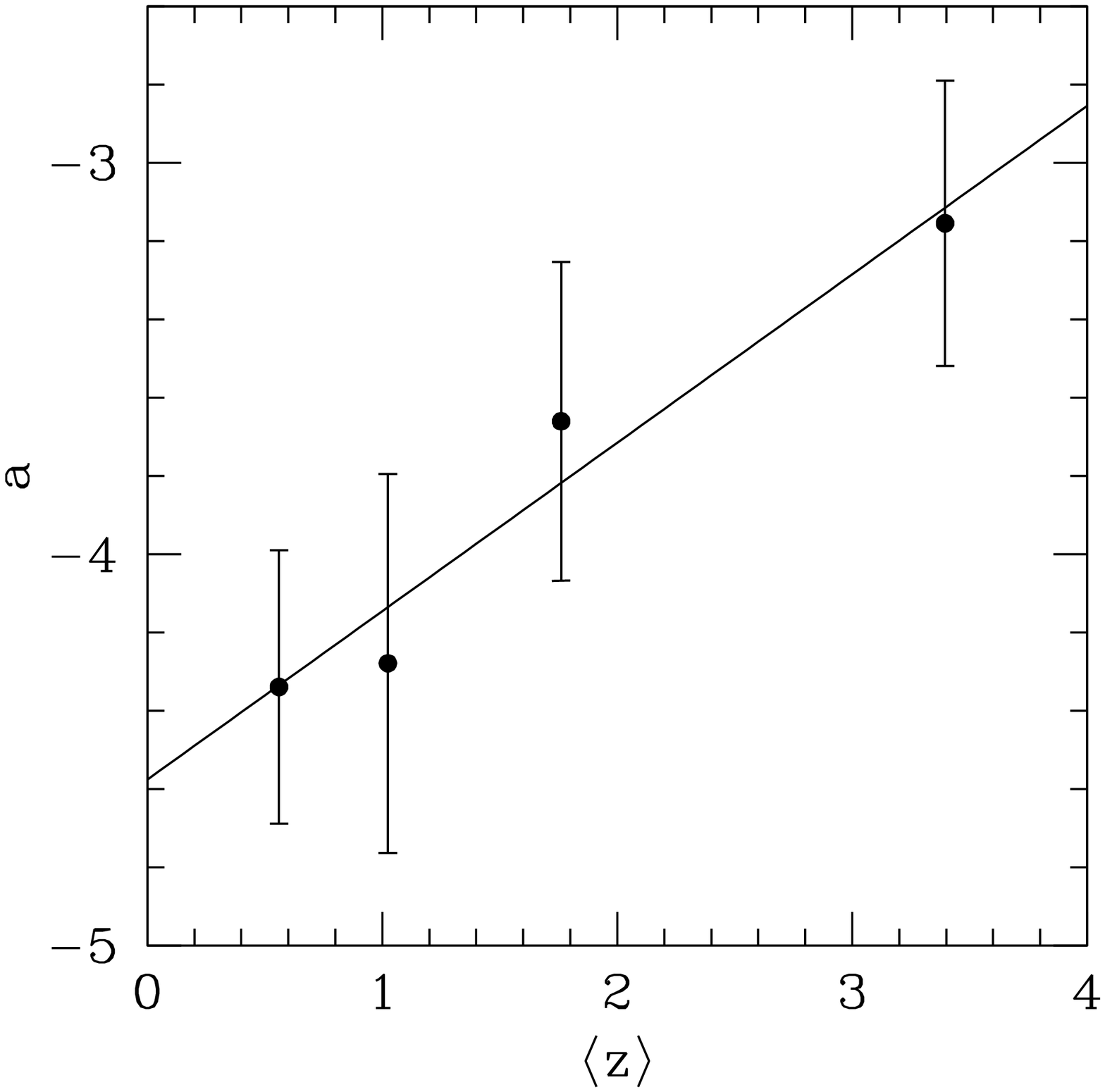}\hspace{0.7cm}
\includegraphics[angle=0,scale=0.35]{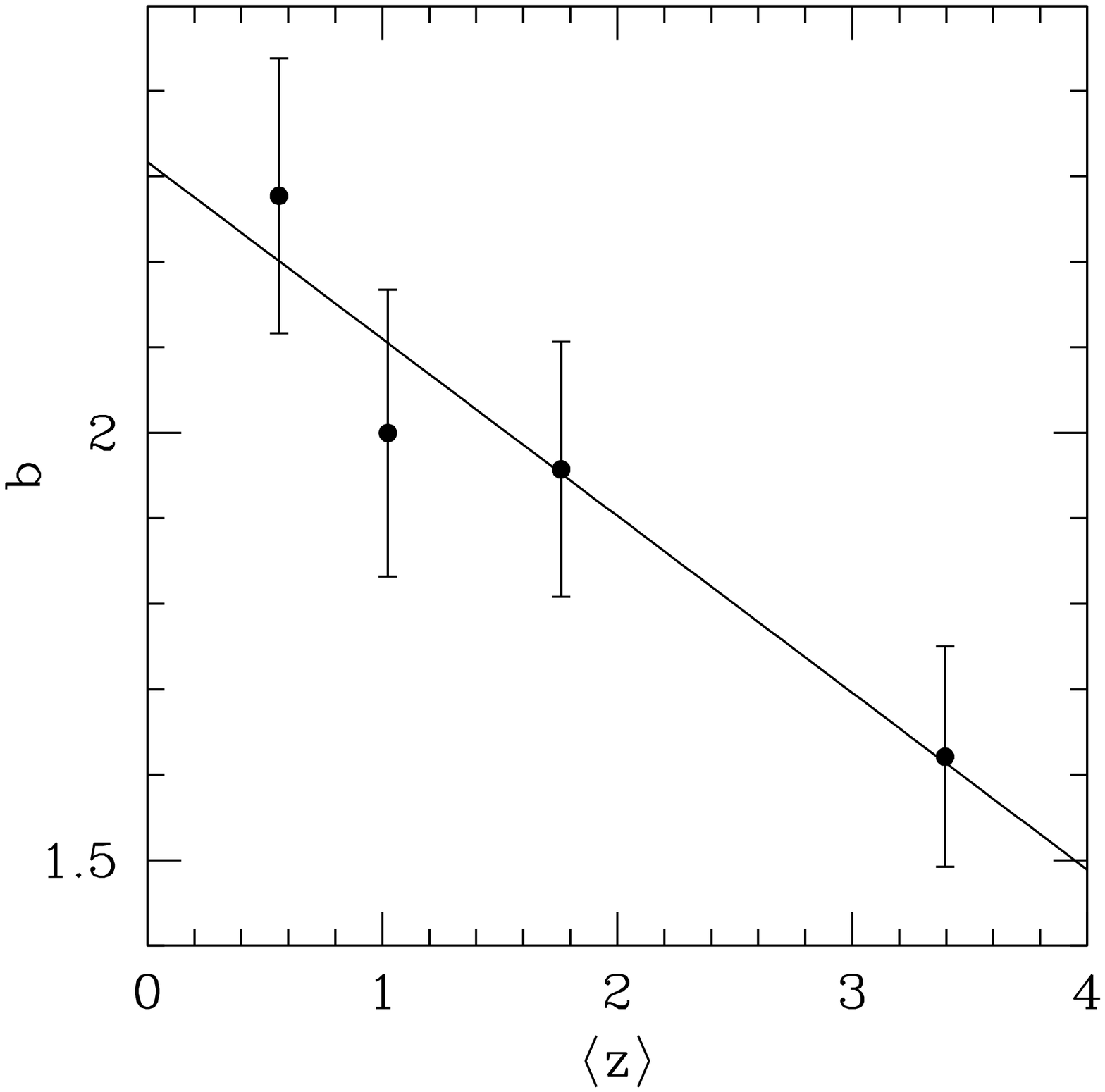}
\caption{The fitted value of $a$ (left panel) and $b$ (right panel) against 
the mean redshift of GRBs. Each data point with error bars represents a group 
of GRBs. The solid line is a least squares linear fit to $a$--$\langle 
z\rangle$ and $b$--$\langle z\rangle$.
}
\label{ee_ab}
\end{figure}

I find that the values of $a$ and $b$ vary with the mean redshift of the 
GRBs monotonically. In Fig.~\ref{ee_ab}, I 
plot $a$ and $b$ against $\langle z\rangle$. Clearly, $a$ and $b$ are 
correlated/anti-correlated with $\langle z\rangle$. The Pearson linear 
correlation coefficient between $a$ and $\langle z\rangle$ is $r(a,\langle 
z\rangle) = 0.975$, corresponding to a probability $P = 0.025$ for a zero 
correlation. The correlation coefficient between $b$ and $\langle z\rangle$ 
is $r(b,\langle z\rangle) = -0.960$, corresponding to $P = 0.040$ for a zero 
correlation.

A least squares linear fit to $a$--$\langle z\rangle$ leads to
\begin{eqnarray}
	a = -4.58 + 0.43 \, z \;, \label{az_f}
\end{eqnarray}
with $\chi^2_r = 0.13$. A least squares linear fit to $b$--$\langle z\rangle$
leads to
\begin{eqnarray}
	b = 2.32 - 0.207 \, z \;, \label{bz_f}
\end{eqnarray}
with $\chi^2_r = 0.31$.

The results indicate that $a$ and $b$ strongly evolve with the cosmic
redshift.

I have used Monte-Carlo simulations to test if the variation of $a$ and $b$ 
with the redshift is caused by the selection effect. The results
show that there is only a $\sim 4$ percent chance that the observed variation 
is caused by the selection effect arising from the limit in the GRB fluence
\citep{li07}. Hence, the variation of the Amati relation with redshift that 
I have discovered may reflect the cosmic evolution of GRBs and indicates that 
GRBs are not standard candles.

The results need to be tested with more GRBs. No matter what the conclusion 
will be (whether the variation of parameters is caused by the GRB evolution 
or by the selection effect), the results suggest that it is very risky to use 
GRBs with redshifts spanning a large range as a single sample to analyze 
physical relations among observables. Although I have only tested the Amati 
relation, it would be surprising if any of the other relations does not 
change with redshift.

\end{document}